# Anomalous length dependent conductance of quasi-one-dimensional molecular wires assembled from metal superatoms

Famin Yu[1,2,3], Rui-Qin Zhang*[4] and Zhigang Wang*[2,3]

[1] School of Physics, Changchun Normal University, Changchun 130032, China

[2] Key Laboratory of Material Simulation Methods & Software of Ministry of Education, College of Physics, Jilin University, Changchun 130012, China

[3] Institute of Atomic and Molecular Physics, Jilin University, Changchun 130012, China

[4] Department of Physics, City University of Hong Kong, Kowloon 999077 Hong Kong SAR, China

E-mail: aprqz@cityu.edu.hk (R. Z.); wangzg@jlu.edu.cn (Z. W.)

**Abstract**

Molecular wires with high electrical conductance are desirable components for future molecular-scale circuitry. However, their conductance typically decays exponentially with increasing length. Here, we report a novel discovery that the conductivity of a nanoscale molecular wire assembled from metal superatoms increases with length. Specifically, high-precision first-principles calculations show that, while the conductance of quasi-one-dimensional superatomic assemblies formed with individual W@Cu$_{12}$ superatoms as units exhibits a slow decay with increasing length, by extending the quasi-one-dimensional superatomic assemblies into bundle-like structures, their electrical conductivity increases with length, accompanied by a change in the corresponding decay factor from 1.25 nm$^{−1}$ to -0.95 nm$^{-1}$. This significant change in the decay factor originates from that the Fermi level of the superatomic assemblies shifts closer to the Fermi level of the electrodes, thereby reducing the tunneling barrier. Our findings highlight that molecular-scale devices exhibiting a negative decay factor are not an inherent characteristic, but rather can be regulated through structural modulation. This work not only provides theoretical references for modulating the performance of molecular-scale circuitry, but also opens up the application prospect of metal superatoms in electrical transport field.

**Introduction**

The development of electronics is trending towards the molecular-scale, requiring the understanding of material structures at the atomic level[1-4]. One of the important researches is to discover molecular wires (MWs) that facilitate efficient charge transport over long distances [5-7]. However, numerous studies have shown that the charge transport mechanism for nanoscale MWs shorter than ~2.5 nm is attributed to tunnelling [5, 8-10], and the conductance $G_0$ decays exponentially by molecular length L according to equation $G_0 = A*\exp(-\beta L)$, where A is the prefactor and $\beta$ is the decay factor [11, 12]. In this context, the search for small values of β in molecular-scale electronics has been continuing for decades [11, 13-15]. Until 2018, a theoretical study reported fuse-assembled quasi-1D porphyrin-based MWs possess vanishing or even a negative β value [16]. Subsequently, in 2020, a research based on scanning tunneling microscope based-break junction (STM-BJ) technique indicated that cumulene wires display increasing conductance with increasing length [17]. Moreover, the research based on comparative analysis methods reveals that this abnormal electrical transport phenomenon occurs in assemblies where the units are strongly coupled [16, 17]. Thus, the quest for molecular-scale devices with increasing conductance with increasing length can begin with the search for units assembled by strong coupling interactions.

Superatoms are molecular systems composed of multiple atoms whose orbitals exhibit angular momentum symmetry similar to that of electrons arranged in shells within atoms [18-20]. As fundamental building blocks at the atomic level, they enable the bottom-up assembly of a wide range of functional systems, thereby facilitating the exploration of unique and uncommon physical phenomena [21-28]. For instance, the asymmetric monolayer assembled by $C_{60}$ superatoms endows the anisotropic phonon modes and conductivity, forming a topological structure that distinguishes them from traditional two-dimensional materials [29, 30]. Endohedral metallofullerene superatoms can achieve bottom-up customized assembly through spin-polarized magnetic coupling, such as superatomic-based chiral assemblies [31-33]. It is noteworthy that recent studies have reported that $W@Cu_{12}$ superatoms can be directly assembled into oligomers and crystals through highly conjugated superatomic molecular orbitals (SAMOs) [34, 35], which may directly benefit electrical transport [11, 16, 36]. In addition, considering that the metal systems generally exhibit better conductivity than many organic molecules [5, 37], it can be inferred that highly conjugated superatomic molecular wires (SMWs) assembled from $W@Cu_{12}$ metal superatoms have the potential to exhibit enhanced conductance with increasing length.

In this work, we conducted research on the electrical transport properties of quasi-1D assemblies using superatoms as building blocks. Specifically, we selected the structurally stable $W@Cu_{12}$ with a closed-shell electronic configuration as units and performed this study using the non-equilibrium Green's function (NEGF) formalism within the density functional theory (DFT). The results show that the decay factor of quasi-1D SMWs is smaller than that of many organic molecular junctions, and that quasi-1D bundle-like SMWs exhibit enhanced conductance with increasing length. Projected density of states (PDOS) analysis reveals that the enhancement of conductance with increasing length is attributed to the gradual convergence of SMWs' Fermi level toward that of the electrodes. This study contributes to the development of superatomic-based high conductance molecular-scale devices.

**Methods**

The $W@Cu_{12}$ quasi-1D SMWs were optimized using density functional theory (DFT) in Dmol$^3$ software. The Perdew-Burke-Ernzerhof (PBE) functional within Generalized Gradient Approximation (GGA) [38] was used to treat the exchange-correlation functional [39] and the Double Numerical plus polarization (DNP) basis set was applied. To calculate the electrical transport properties of quasi-1D $W@Cu_{12}$ SMWs, we adopted the non-equilibrium Green's function (NEGF) formalism within the DFT implemented by the Nanodcal software [40]. In the NEGF- DFT calculation, GGA-PBE96 was used for the exchange-correlation functional, and the double-ζ plus polarization (DZP) basis set was used for all atoms. We used a two-probe model which comprises three parts: left and right electrodes (which extend to ± ∞) plus central scattering region. The distance between the central scattering region and the electrodes was maintained at 2.5 Å for different quasi-1D W@Cu$_{12}$ SMWs to minimize differences in the electrode-scattering region coupling, which contributes to a clearer understanding of the physics arising from the device length. Vacuum layers in the y and z directions exceeding 10 Å were adopted to eliminate interactions between neighboring cells. The cutoff energy was set to 80 Hartree, the temperature of electrodes was chosen as 300 K and the k-point grid of electrodes was set to 1×1×100.

## Results and discussion

To investigate the electrical transport properties of metal SMWs, the highly conjugated quasi-1D W@Cu$_{12}$ SMWs were constructed, where the W atom is positioned at the core of a three-dimensional cage structured by an arrangement of 12 Cu atoms, as shown in Figure 1(a). The most stable spontaneous assembly approach of W@Cu$_{12}$ is rotating the superatoms separately 45 degrees perpendicular to the axis in the assembly direction [34, 35, 41], thereby forming quasi-1D W@Cu$_{12}$ SMWs. Typically, the mechanism governing charge transport in linear structures shorter than approximately 2.5 nm is tunnelling, and the quantum-mechanically governed tunnelling effect is a key factor in the appearance of an increase in conductance with length. Thus, the assembly of the W@Cu$_{12}$ superatom ends after the addition of four units, resulting in quasi-1D SMWs with lengths of 0.54, 1.11, 1.70, and 2.28 nm, respectively.

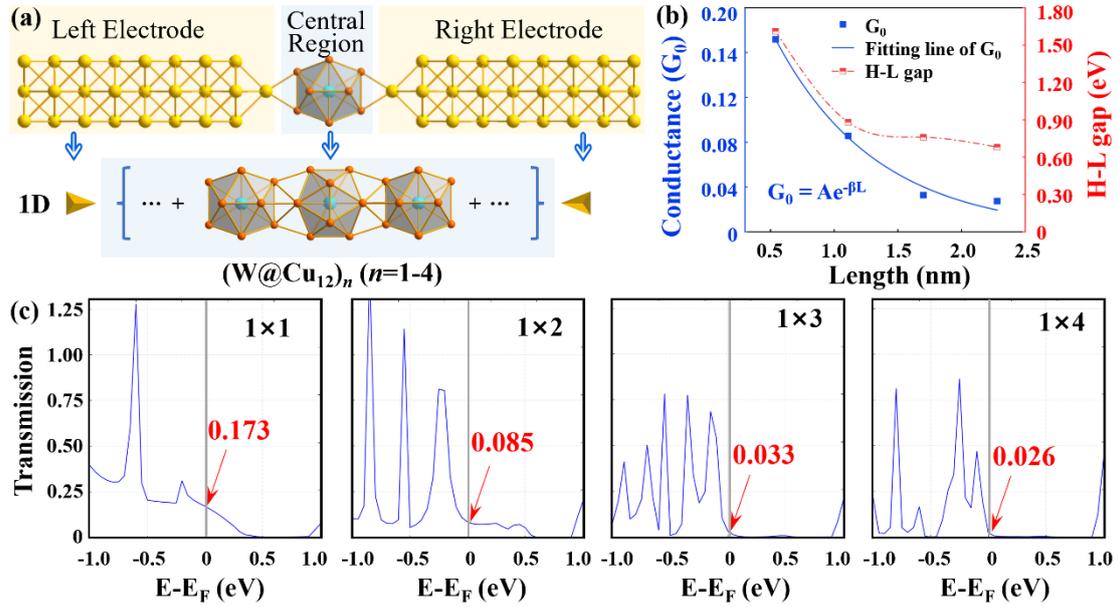

**Figure 1.** Models and electrical transport properties of quasi-1D W@Cu$_{12}$ SMWs. (a) Two-probe structures are used in the electrical transport calculations. The positions of Au electrodes are indicated by yellow triangles. (b) Electrical conductance and HOMO-LUMO gap of quasi-1D W@Cu$_{12}$ SMWs with different lengths. H and L indicate HOMO and LUMO, respectively. (c) Average transmission spectra for quasi-1D W@Cu$_{12}$ SMWs with different lengths. The corresponding Fermi level is indicated by the gray line, and the transmission function at the Fermi level is marked by the red label.

To calculate the electrical conductance of quasi-1D W@Cu$_{12}$ SMWs, the two-probe devices using Au probes as semi-infinite electrodes were employed to connect the central scattering region. As shown in Figure 1(a), quasi-1D W@Cu$_{12}$ SMWs in the central scattering region are laterally connected to electrodes, and one Au atom at each electrode tip is connected to two Cu atoms on the quasi-1D SMWs. The first-principles calculations show that the conductance of 1×1, 1×2, 1×3, and 1×4 W@Cu$_{12}$ quasi-1D W@Cu$_{12}$ SMWs are 0.172, 0.085, 0.033 and 0.027 G$_0$, respectively. The decrease in conductance with the length is fitted in Figure 1(b) and the decay factor of conductance is determined to be 1.25 nm$^{-1}$, while the decay factor ranges from 2-6 nm$^{-1}$ for conjugated molecules [8, 42-44], and 6-12 nm$^{-1}$ for nonconjugated molecules [45-47]. Such a low decay constant facilitates quasi-1D W@Cu$_{12}$ SMWs to become a promising candidate for interconnect in molecular-scale circuitry.

To explore the possibility of conductance increase with length, it is imperative to first understand why the conductance of quasi-1D W@Cu$_{12}$ SMWs diminishes slightly as their length increases. Here, the highest occupied molecular orbitals (HOMO) - the lowest unoccupied molecular orbitals (LUMO) gap was analyzed, which influences the conductance [48, 49]. As shown in Figure 1(b), the HOMO-LUMO gap exhibits dramatic narrowing with increasing length, leading to a slow reduction in the conductance. In addition, the transmissions represent the probability that a charge with a given energy will transmit through a system between the electrodes. As shown in Figure 1(c), the transmission function at the Fermi level for W@Cu$_{12}$, 2W@Cu$_{12}$, 3W@Cu$_{12}$ and 4W@Cu$_{12}$ SMWs is 0.173, 0.085, 0.033 and 0.026, respectively. The relationship between the transmission coefficient and conductance is given by [50]:

$$G_0 = \frac{e^2}{h} T(E_F)$$

where e is the electron charge, $h$ is Planck's constant and $E_F$ is the Fermi level. The transmission function decreases slowly as the length of 1D W@Cu$_{12}$ increases, which agrees with the conductance trend. The discoveries here present promising prospects for achieving conductance regulation.

Considering the factors affecting the molecular orbitals and subsequently the gap, we directly choose the approach of increasing the radius in the perpendicular 1D direction, that is, forming bundle-like structures. The reason for this choice is that increasing the units in the perpendicular 1D direction can introduce strong coupling effects in a new dimension [34], which is generally recognized as potentially having a significant impact on the electronic structure [51, 52]. Here, it should be noted that when the length of the 1×1 structure is fixed, expanding it into a bundle-like configuration would lead to its transformation into a monolayer arrangement, which is therefore not considered in the present work. Thus, the 2×2×2 W@Cu$_{12}$ octamer is regarded as an initial system, and the length is increased layer by layer until it is increased to a 2×2×4 16-polymer, as shown in Figure 2(a)-(b). The lengths of these quasi-1D bundle-like W@Cu$_{12}$ SMWs are 1.14, 1.73 and 2.30 nm, respectively. In this way, the device length is below 2.5 nm, ensuring that tunnel transport dominates.

To reliably calculate the conductance of quasi-1D bundle-like W@Cu$_{12}$ SMWs and effectively compare the results with different radii, the same two-electrode configuration as quasi-1D W@Cu$_{12}$ SMWs was adopted (Figures 2(a)-(b)). The results show that the conductance for 2×2×2, 2×2×3, and 2×2×4 W@Cu$_{12}$ SMWs is 0.105, 0.117 and 0.253 G$_0$, respectively. From Figure 2(c), the conductance increases with the length and the decay constant is determined to be -0.95 nm$^{-1}$. Especially, the conductance of 2×2×4 SMWs even surpasses that of individual W@$Cu_{12}$ monomer. Additionally, the calculated transmission functions at the Fermi level for 2×2×2, 2×2×3 and 2×2×4 SMWs are 0.047, 0.055 and 0.069, respectively. This upward trend is aligned with the conductance variability, further corroborating the reliability of the conductance trend. The negative attenuation of quasi-1D bundle-like W@$Cu_{12}$ SMWs renders them candidates for the construction of molecular electronic devices.

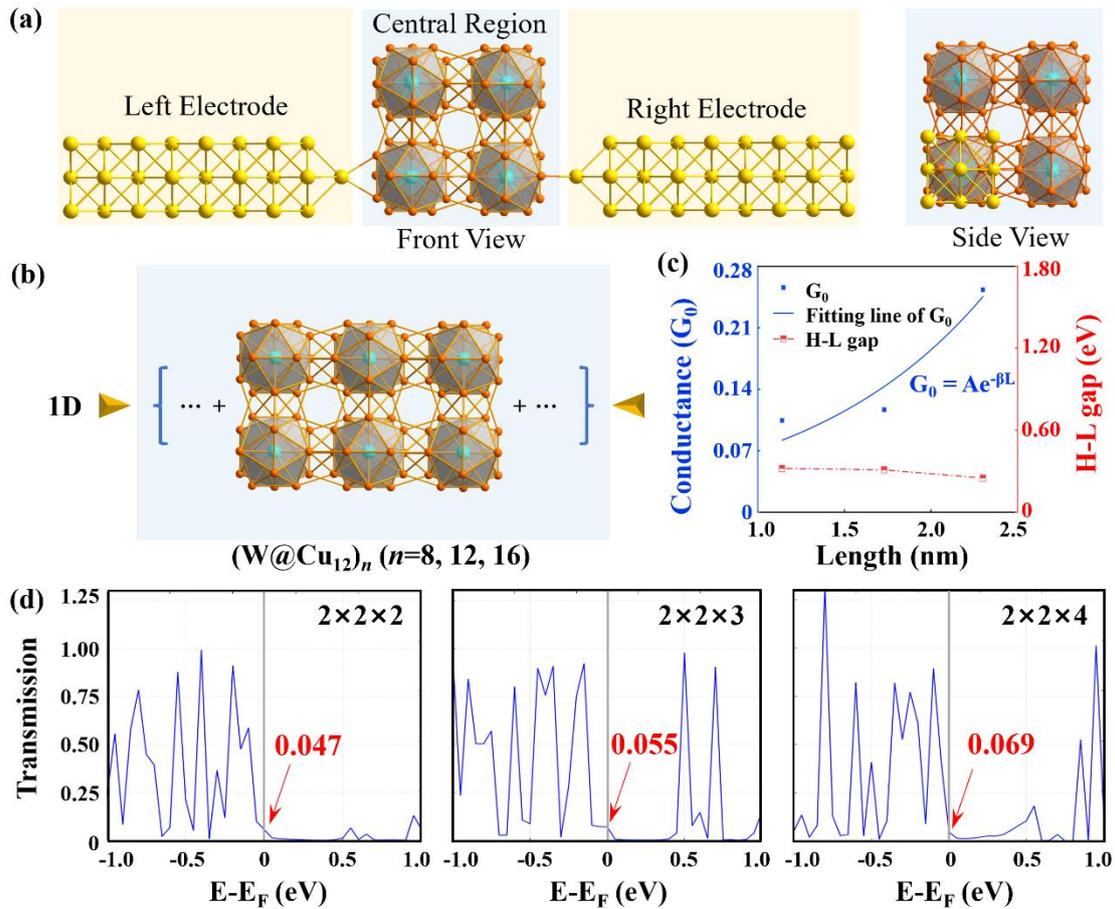

**Figure 2.** Models and electrical transport properties of quasi-1D bundle-like W@$Cu_{12}$ SMWs. (a)-(b) Two-probe structures are used in the electron transport calculations. The positions of Au electrodes are indicated by yellow triangles. (c) Electrical conductance and HOMO-LUMO gap of quasi-1D bundle-like W@$Cu_{12}$ SMWs with different lengths. H and L indicate HOMO and LUMO, respectively. (d) Average transmission spectra for quasi-1D bundle-like W@$Cu_{12}$ SMWs with different lengths. The corresponding Fermi level is indicated by the gray line, and the transmission function at the Fermi level is marked by the red label.

It is noteworthy that as the size of quasi-1D bundle-like W@$Cu_{12}$ SMWs increases from 2×2×2 to 2×2×4, their HOMO-LUMO gaps decrease from 0.32 eV to 0.25 eV, showing a downward trend of $10^{-1}$ eV. This small variation is insufficient to directly explain the enhancement of the conductance with increasing length. To gain deeper insight into the uncommon transport properties emerging upon radius expansion, the PDOS analysis was performed to investigate the relative positioning of HOMO, LUMO and Fermi levels within the central scattering region. Among them, the energy of HOMO and LUMO can be determined by the peak position of PDOS, and the Fermi level lies somewhere between them. As shown in Figure 3(a), for quasi-1D W@$Cu_{12}$ SMWs, an increase in length leads to an upward energy shift of HOMO, while the LUMO remains relatively stable. This shift results in a rightward shift of the Fermi level, which in turn enlarges the distance between the Fermi level of the device and the electrodes. Conversely, Figure 3(b) shows that as the quasi-1D W@$Cu_{12}$ SMWs extend to bundle-like configuration, both the HOMO and LUMO undergo an energy reduction with increasing device length. This approaching effect reduces charge tunnel barriers [16], elucidating why the conductance of quasi-1D bundle-like W@$Cu_{12}$ SMWs enhances with increasing length.

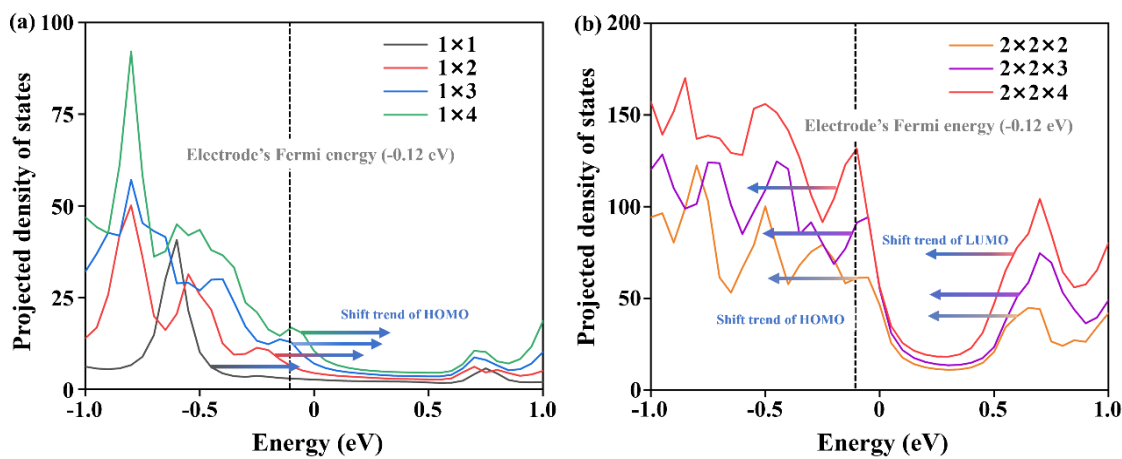

**Figure 3.** Projected density of states (PDOS) of SMWs. (a) PDOS of quasi-1D W@$Cu_{12}$ SMWs. (b) PDOS of quasi-1D bundle-like W@$Cu_{12}$ SMWs. The gray dashed line represents the Fermi level of the electrode, as determined by a self-consistent energy calculation specific to the electrode. The arrows indicate the tendency of HOMO and LUMO to shift in energy as the length of the SMWs increases.

## Conclusion

In summary, by using typical superatom W@$Cu_{12}$ as building blocks to construct the highly conjugated quasi-1D SMWs, we investigated the electron transport properties of quasi-1D and quasi-1D bundle-like SMWs. First-principles calculations reveal that the decay factor of quasi-1D W@$Cu_{12}$ SMWs is determined to be 1.25 $nm^{-1}$, which is smaller than that of many organic MWs. More importantly, the conductance of quasi-1D bundle-like W@$Cu_{12}$ SMWs is enhanced with increasing length. Besides, the 2×2×4 quasi-1D bundle-like W@$Cu_{12}$ SMWs even exhibit higher conductance than W@$Cu_{12}$ monomer. This trend of increasing conductance with SMWs length can be attributed to structural modulation that bring the Fermi level of the quasi-1D bundle-like W@$Cu_{12}$ SMWs closer to that of the electrode, thereby reducing the tunnel barriers. Therefore, this study contributes to the regulation of molecular-scale circuit performance and promotes the application of superatomic-based electronic devices.

## Conflict of interest

The authors declare no conflict of interest.

## Acknowledgements

This work is supported by the National Natural Science Foundation of China (under grant numbers 11974136) and the Research Grant Council of Hong Kong SAR (11317122). We gratefully acknowledge HZWTECH for providing computation facilities. Z. W. also acknowledges the High-Performance Computing Center of Jilin University and National Supercomputing Center in Shanghai.**References**

[1] Liu, J., X. Zhao, J. Zheng, X. Huang, Y. Tang, F. Wang, R. Li, J. Pi, C. Huang, L. Wang, Y. Yang, J. Shi, B.-W. Mao, Z.-Q. Tian, M.R. Bryce, and W. Hong 2019 Transition from Tunneling Leakage Current to Molecular Tunneling in Single-Molecule Junctions *Chem* **5** 390-401

[2] Khan, H.N., D.A. Hounshell, and E.R.H. Fuchs 2018 Science and research policy at the end of Moore's law *Nat. Electron.* **1** 14-21

[3] Xiang, D., X. Wang, C. Jia, T. Lee, and X. Guo 2016 Molecular-Scale Electronics: From Concept to Function *Chem. Rev.* **116** 4318-440

[4] Joachim, C., J.K. Gimzewski, and A. Aviram 2000 Electronics using hybrid-molecular and mono-molecular devices *Nature* **408** 541-8

[5] Feng, A., S. Hou, J. Yan, Q. Wu, Y. Tang, Y. Yang, J. Shi, Z.-Y. Xiao, C.J. Lambert, N. Zheng, and W. Hong 2022 Conductance Growth of Single-Cluster Junctions with Increasing Sizes *J. Am. Chem. Soc.* **144** 15680-8

[6] Vilan, A., D. Aswal, and D. Cahen 2017 Large-Area, Ensemble Molecular Electronics: Motivation and Challenges *Chem. Rev.* **117** 4248-86

[7] Tan, Z., D. Zhang, H.-R. Tian, Q. Wu, S. Hou, J. Pi, H. Sadeghi, Z. Tang, Y. Yang, J. Liu, Y.-Z. Tan, Z.-B. Chen, J. Shi, Z. Xiao, C. Lambert, S.-Y. Xie, and W. Hong 2019 Atomically defined angstrom-scale all-carbon junctions *Nat. Commun.* **10** 1748

[8] Ho Choi, S., B. Kim, and C.D. Frisbie 2008 Electrical Resistance of Long Conjugated Molecular Wires *Science* **320** 1482-6

[9] Davis, W.B., W.A. Svec, M.A. Ratner, and M.R. Wasielewski 1998 Molecular-wire behaviour in p-phenylenevinylene oligomers *Nature* **396** 60-3

[10] Hines, T., I. Diez-Perez, J. Hihath, H. Liu, Z.-S. Wang, J. Zhao, G. Zhou, K. Müllen, and N. Tao 2010 Transition from Tunneling to Hopping in Single Molecular Junctions by Measuring Length and Temperature Dependence *J. Am. Chem. Soc.* **132** 11658-64

[11] Leary, E., B. Limburg, A. Alanazy, S. Sangtarash, I. Grace, K. Swada, L.J. Esdaile, M. Noori, M.T. González, G. Rubio-Bollinger, H. Sadeghi, A. Hodgson, N. Agraït, S.J. Higgins, C.J. Lambert, H.L. Anderson, and R.J. Nichols 2018 Bias-Driven Conductance Increase with Length in Porphyrin Tapes *J. Am. Chem. Soc.* **140** 12877-83

[12] Sangtarash, S., A. Vezzoli, H. Sadeghi, N. Ferri, H.M. O'Brien, I. Grace, L. Bouffier, S.J. Higgins, R.J. Nichols, and C.J. Lambert 2018 Gateway state-mediated, long-range tunnelling in molecular wires *Nanoscale* **10** 3060-7

[13] Sadeghi, H., S. Sangtarash, and C. Lambert 2017 Robust Molecular Anchoring to Graphene Electrodes *Nano Lett.* **17** 4611-8

[14] Sedghi, G., K. Sawada, L.J. Esdaile, M. Hoffmann, H.L. Anderson, D. Bethell, W. Haiss, S.J. Higgins, and R.J. Nichols 2008 Single Molecule Conductance of Porphyrin Wires with Ultralow Attenuation *J. Am. Chem. Soc.* **130** 8582-3

[15] Fan, F.-R.F., J. Yang, L. Cai, D.W. Price, Jr., S.M. Dirk, D.V. Kosynkin, Y. Yao, A.M. Rawlett, J.M. Tour, and A.J. Bard 2002 Charge Transport through Self-Assembled Monolayers of Compounds of Interest in Molecular Electronics *J. Am. Chem. Soc.* **124** 5550-60

[16] Algethami, N., H. Sadeghi, S. Sangtarash, and C.J. Lambert 2018 The Conductance of